# Time-reversal and super-resolving phase measurements

K. J. Resch[1], K. L. Pregnell[1,2], R. Prevedel[1], A. Gilchrist[1,2], G. J. Pryde[1,2], J. L. O'Brien[1,2] and A. G. White[1,2]

[1]*Department of Physics and* [2]*Centre for Quantum Computer Technology,*
*University of Queensland, Brisbane QLD 4072, Australia*

We demonstrate phase super-resolution in the absence of entangled states. The key insight is to use the inherent time-reversal symmetry of quantum mechanics: our theory shows that it is possible to *measure*, as opposed to prepare, entangled states. Our approach is robust, requiring only photons that exhibit classical interference: we experimentally demonstrate high-visibility phase super-resolution with three, four, and six photons using a standard laser and photon counters. Our six-photon experiment demonstrates the best phase super-resolution yet reported with high visibility and resolution.

Common wisdom holds that entangled states are a necessary resource for many protocols in quantum information. An example is quantum metrology, which promises super-precise measurement, surpassing that possible with classical states of light and matter [1, 2]. In the last 20 years quantum metrology schemes have been proposed for improved optical [3–8] and matter-wave [9] interferometry, atomic spectroscopy [10], and lithography [11–13]. The entangled states in these schemes give rise to *phase super-resolution*, where the interference oscillation occurs over a phase N-times smaller than one cycle of classical light [14, 15] and *phase super-sensitivity*, a reduction of phase uncertainty.

Many quantum metrology schemes are based on path-entangled number states. The canonical example is the NOON-state [1], a two-mode state with either N particles in one mode and 0 in the other or vice-versa, i.e., $(|N0\rangle+|0N\rangle)/\sqrt{2}$. A deterministic optical source of path-entangled states is yet to be realised, requiring optical nonlinearities many orders of magnitude larger than those currently possible. However, entangled states can be made *non-deterministically* using single-photon sources, linear optics, and photon-resolving detectors [16]: leading to a flurry of proposals to generate path-entangled states [17–20]. While phase super-resolution with two-photons has been demonstrated often since 1990 [21–24], phase super-resolution was experimentally demonstrated for 3-photon [14] and 4-photon [15] states only recently. As efficient photon sources and photon-number resolving detectors do not yet exist, all demonstrations to date necessarily used multiphoton coincidence post-selection [25]. Problematically, current photon sources are extremely dim and true photon-number resolving detectors are expensive and uncommon. In this paper we introduce a time-reversal technique that eliminates the need for exotic sources and detectors, achieving high-visibility phase super-resolution with a standard laser and photon detectors.

Fig. 1a) depicts a method for probabilistically generating NOON states via linear optics and post-selection. Single photon states are prepared in each of the N input modes, $|\Psi_i\rangle=|11...1\rangle_{12...N}$, of a linear optical multi-port interferometer, $U_{\text{multi}}$ [26]. With probability $\eta_p$, no photons are found in modes 3 to N, heralding the NOON-state in modes 1 and 2, $(|N0\rangle_{12}+|0N\rangle_{12})/\sqrt{2}$. A relative phase shift, $\phi$, between modes 1 & 2 introduces a N$\phi$ shift between the terms in the state—phase super-resolution. Maximum fringe visibility will be achieved when the system is measured in a state $\langle\psi_N|$ which has equal overlap, $\kappa_N=|\langle\psi_N|N0\rangle|^2=|\langle\psi_N|0N\rangle|^2$, with both components of the NOON-state. Fig. 1 shows two possible measurements, i) yields $\kappa_N^i=1/2^N$ [19], ii) yields $\kappa_N^{ii}=\kappa_N^i/\sqrt{2\pi N}$ in the optimum case $|\alpha|^2=N/2$. The probability of detecting a final state $\langle\Psi_f|=\langle\psi_N|_{12}\langle0...0|_{3...N}$ after propagating through the multiport and phase shifter, $U_\phi$, is $P=|\langle\Psi_f|U_\phi U_{\text{multi}}|\Psi_i\rangle|^2=\eta_p\kappa_N(1+\cos N\phi)$. This probability exhibits phase super-resolution since the fringes complete N oscillations over a single cycle of $2\pi$.

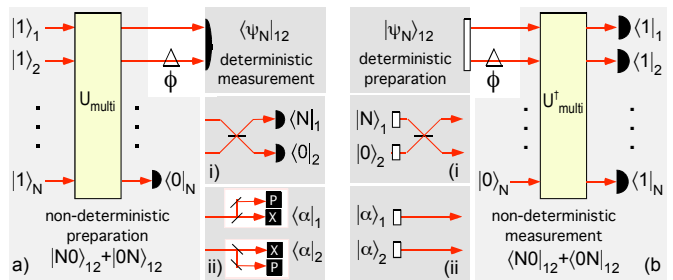

FIG. 1: Non-deterministic a) preparation & (b measurement of NOON-states for phase super-resolution, as described in text. i) photon-counting after a 50% beamsplitter to measure $\langle N|_1$ & $\langle 0|_2$. ii) coherent-state detection, $\langle \alpha|_1$ & $\langle \alpha|_2$, via a 50% beamsplitter and two homodyne detectors to measure amplitude and phase. (i & (ii are the corresponding time-reversed processes.

Probabilities in quantum mechanics are invariant under time reversal [27–29], i.e., if we swap the input and measured states and suitably time-reverse the operation of the multiport, as shown in Fig. 1(b, the probability is unchanged, $P=|\langle\Psi_i|U^\dagger_{\text{multi}}U^\dagger_\phi|\Psi_f\rangle|^2=\eta_p\kappa_N(1+\cos N\phi)$. In the time-reversed picture, the interferometer no longer plays the role of probabilistic NOON-state generator, but rather constitutes a probabilistic NOON-state *detector*: since the probability, $P$, is invariant under time reversal, phase super-resolution is also invariant. Experimentally, detecting NOON-states is much easier than creating them: time-reversing turns the difficult generation of N single photons into straightforward detection of N photons in coincidence, and turns the problematic detection of the vacuum into vacuum inputs which are automatically available with perfect fidelity. Successful detection of a NOON-state is signalled by the coincident detection of a single photon at each output detector; this is the time-reverse of the coincident creation of a single photon at each input. This time-reversal technique is a simple example of a more general measurement technique introduced by Pregnell and Pegg [29, 30].

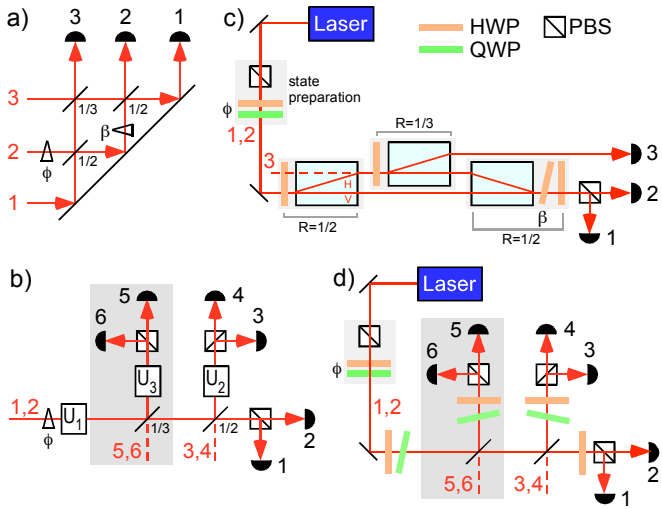

FIG. 2: Ideal multiports, $U_{\rm multi}$: a) symmetric 3×3 and b) asymmetric 4×4 (6×6), constructed from 2×2 beamsplitters with reflectivities as shown (PBS, polarising beamsplitter). Laser light is input to modes 1, 2, with no light to modes 3-6. In a) the internal phase, $\beta$, ensures each input transforms to an equal superposition of the three outputs; in b) polarisation rotations $U_1$ to $U_3$ set the phases of the singles fringes. c), d) Experimental realisations of a), b). In c) the indicated HWP's are set to 22.5° to form 1/2 beamsplitters with the beam displacers, the third is set to 17.6° so that 2/3 of the light intensity takes the upper path; the angle of the tilted HWP sets $\beta$, its optic axis is at 0°. In d), the beamsplitter for modes 3, 4 is a pellicle; for modes 5, 6 it is a microscope slide set at a small angle of incidence, ∼10°, to avoid polarisation effects. Reflectivities are not ideal, the rates are equalised with lossy coupling. $U_1$ to $U_3$, are realised using waveplates: the orientation and tilt of each waveplate was adjusted so that one detector reaches an interference minimum every 22.5° for N=4; every 15° for N=6.

This theory assumes that the input photons are indistinguishable. A significant advantage of our approach is that it is robust: phase super-resolution occurs even when the photons are distinguishable. Creating NOON states relies on non-classical interference, which requires indistinguishable photons. In our time-reversed scheme we require *only* that the inputs display classical interference: this is not affected by the degree of distinguishability *between* the photons, or the photon arrival statistics. Experimentally, there is a trade-off between temporal distinguishability and counting rate: photons become distinguishable as the coincidence window time is increased above the input light coherence time, but this increases the counting rate. We run in the high counting rate limit to achieve the best statistics, limited only by saturation effects in our coincidence-counting electronics.

In our experiments the two bright inputs to the multiport, modes 1 & 2, are the vertical and horizontal polarisation modes of the one spatial mode from a laser. We use an attenuated He:Ne laser (Uniphase 1135P) and set the polarisation with a half-wave plate (HWP) followed by a quarter-wave plate (QWP) at an angle of 45°. Changing the angle of the HWP by $\phi/4$ changes the relative phase between the modes by $\phi$, while ensuring the vertical & horizontal modes are the same amplitude. In classical interferometry, this yields one oscillation for $0<\phi<2\pi$.

Multiports can be symmetric (every input mode is converted into an equal superposition of N output modes [19]) or asymmetric (not every input satisfies this condition [18]). Scaling up symmetric multiports beyond N=2 can be done either with a polynomial number of nested standard interferometers [26], which would be arduous to phase lock, or a single N×N fused fibre except that it is not known how to control the large set of internal phases [19]. Fortunately symmetric multiports are not required for phase super-resolution: an asymmetric multiport suffices for even-N. Fig. 2 shows our symmetric N=3, and asymmetric N=4, 6 multiports (the N=4 multiport was independently proposed in [31]): all designs are passively stable and do not require active phase-locking.

In fig. 2c) the output modes are sent to three pinhole photon counting detectors, D1–D3, where the small aperture is a single-mode fibre without a coupling lens; in Fig. 2d) each output mode is first passed through a polarising beamsplitter and then detected. The singles rate is the number of photons per second detected by an individual detector: for N=3 the maximum was $5\times 10^4$ Hz; for N=4, 6 the maximum singles rate was $1.3\times 10^5$ Hz. The N singles rates are recorded individually. For N=3, the N-fold coincidence rate is measured using two ORTEC 567 Time-to-Amplitude Converter/Single Channel Analyzer (TAC/SCA) modules each with a $1.5\mu s$ coincidence-window; for N=4, 6 coincidence counting was performed using up to 3 TAC/SCAs and an ORTEC CO4020 Quad Logic Unit. For N=4 (6) all pulse length inputs to the Quad were set to $1.5\mu s$ ($5\mu s$) as were the coincidence-windows on the TAC/SCA. In all cases, due to a restricted number of recording channels, the singles were measured immediately after a coincidence run. To avoid saturation in the coincidence electronics, the mean number of photons per coincidence-window must be $\leq 1$: for N=3, 4, and 6 it was up to 0.07, 0.15, and 0.48.

Fig. 3 shows the coincidence and singles rates for the N=3 symmetric, and the N=4, 6 asymmetric experiments of Fig. 2c) & d). Fig. 3a)-c) shows the three-, four-, and six- fold coincidence rates as a function of the phase, $\phi$, with three, four, and six distinct oscillations within a single phase cycle. This is in contrast to the fringes observed in the singles rates, Fig. 3d)-f), which undergo only a single oscillation over the same range. This is the experimental signature of phase super-resolution. We emphasize that this was achieved without *production* of a path-entangled state, which would have had the signature of flat singles rates over an optical cycle [15, 24].

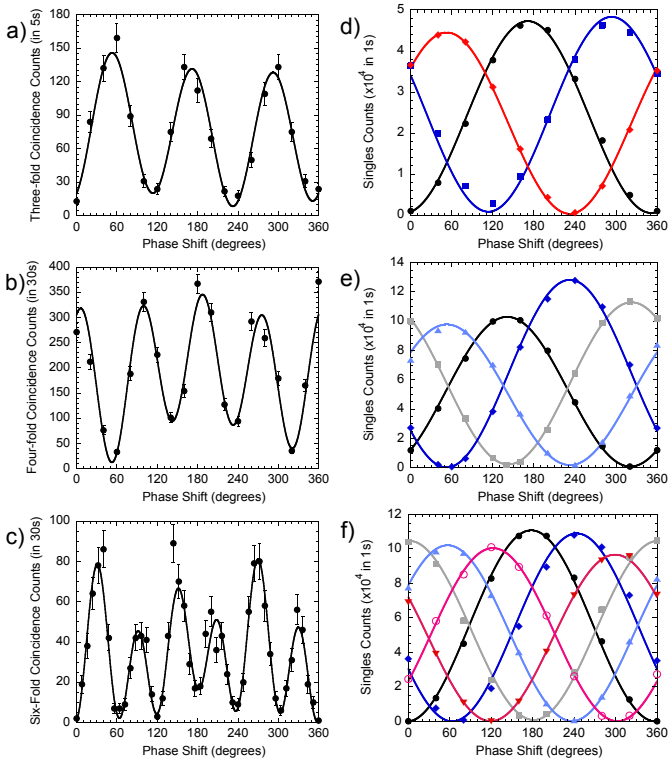

FIG. 3: a)-c) Three-, four- & six- fold coincidence rates as a function of phase, $\phi$, respectively exhibiting 3, 4, and 6 distinct oscillations within a single phase cycle. The main source of uncertainty is Poissonian statistics: error bars represent the square root of the count rate. The solid line is a fit to a product of 3, 4, and 6 sinusoidal fringes, as explained in the text. d)-f) Corresponding singles rates as a function of $\phi$, each exhibiting only one oscillation per phase cycle. Error bars are contained within the data points; solid lines are the individual sinusoidal fringes obtained from a)-c). In d) D1 to D3 are respectively indicated by black, blue, and red; in e)-f) D1 to D6 are indicated by black, grey, blue, cyan, red, and pink. Ideally, the phase differences between adjacent fringes is $2\pi/N$, our fits give: 122° & 119° for N=3; 92°, 90° & 90° for N=4; and 55°, 66°, 56°, 62° & 56° for N=6.

As discussed above, time-reversed phase super-resolution does not rely on non-classical interference: the coincidence rate is determined entirely by the product of the singles rates. Consider an N×N multiport set up so that the detection probability in the $k^{th}$ output mode is $P_k \propto 1+\cos(\phi+2\pi k/N+\varphi)$, where $\varphi$ is a constant phase offset. The N-fold coincidence probability is then simply the product of the single mode probabilities, i.e., $P_{11...1} \propto 1+\cos(N\phi+\Delta(N,\varphi))$ which clearly exhibits N oscillations per cycle, where $\Delta$ is the offset. Applying this to Figs. 3a)-c), we respectively fit a product of 3, 4, and 6 sinusoidal fringes, $s_i=c_i v_i \sin(\phi+\delta_i)+c_i$ where $v_i$ is the visibility, and $c_i$ & $\delta_i$ are amplitude and phase offsets, of the $i^{th}$ fringe. The resulting fits—the solid lines in Figs. 3a)-c)—are very good, with reduced $\chi^2$ of 1.6, 6, and 1.7, respectively. (The high value in the N=4 case is most likely due to observed amplitude instability of the D3 signal during the course of the coincidence measurement.) The coincidence fringes for all three experiments differ from a pure sinusoid due to small variations in the underlying singles rates, and become more pronounced for larger values of $N$. The solid lines in Fig. 3d)-f) are the individual sinusoidal fringes, $s_i$, scaled by a constant factor that matches the amplitude to the data but does not alter the visibility or phase of each fringe. Again the agreement is very good. The N=6 case matches the largest phase super-resolution reported to date, obtained in an ion-trap system [32], but has significantly better visibility.

Our results clearly show that path-entangled states are not required for phase super-resolution [33]. Previous optical demonstrations used nonclassical light sources, which are notoriously dim, limiting the three- and four-fold coincidence rates to 5 Hz [14] and 0.1 Hz [15] respectively [34]. We significantly improve on this, achieving phase super-resolution with a *six*-photon coincidence rate of about 2.7 Hz; furthermore, owing to the high-visibility singles and extremely stable construction of the multi-ports, our fringe patterns all exhibited high visibility. Fitting a single sinusoid, and without any background subtraction, the fringe visibilities for the N=3, 4, and 6 cases are respectively 81±3%, 76±2%, and 90±2%—well exceeding previously reported raw visibilities of 42±3% for N=3 [14] and ∼61% for N=4 [15]. An alternative technique for realising phase super-resolution sums multiple occurrences of a fringe pattern narrowed by nonlinear detection, either spatial [36] or temporal [37]. This suffers from exceedingly low visibility: when the number of exposures equals the number of fringes, $V=2(N!)^2/(2N)!$, for N=6 this predicts $V\sim0.2\%$.

Phase super-*sensitivity* occurs when there is a reduction of the phase uncertainty as compared to that possible with classical resources. Unlike phase super-resolution, phase super-sensitivity cannot be determined solely from the fringe pattern: careful accounting is needed to determine the resource consumption required to achieve the measured signal. For small variations in phase around a given value, the phase uncertainty is $\Delta\phi=\Delta A/\left|\frac{d\langle A\rangle}{d\phi}\right|$, where $A$ and $\Delta A$ are an observable and its associated uncertainty [3]. All other things being equal, the slope in the denominator is increased by phase super-resolution, reducing $\Delta\phi$. Phase super-sensitivity is achieved when $\Delta\phi$ is less than the classical limit,

$$\Delta\phi_{\text{class}} = \frac{1}{\sqrt{N_{tot}}} = \sqrt{\frac{\eta}{N}}, \qquad (1)$$

where $\eta$ allows for non-ideal efficiency in using $N_{tot}$ resources to estimate $\Delta\phi$. Phase super-resolution produces normalized fringes of the form $(1-V\cos N\phi)/2$, where $V$ is the fringe visibility, and the slope is $d\langle A\rangle/d\phi =$

$\frac{1}{2}NV \sin N\phi$. Beating the classical limit requires,

$$\eta V^2 > \frac{4(\Delta A)^2}{N|\sin^2(N\phi)|}. \quad (2)$$

Consider $A$ to be a projector with measurement outcomes bounded by 0 and 1; the worst case is $\Delta A=1/2$. At the point of minimum phase uncertainty, Eq. 2 reduces to,

$$\eta V^2 N > 1. \quad (3)$$

By this criterion, although several experiments have demonstrated phase super-resolution, there has been no unambiguous demonstration of phase super-sensitivity.

The best known preparation efficiency in nondeterministic optical schemes is $\eta=2N!/N^N$ [17–19]. In the ideal limit, Eq. 3 gives $2N!/N^{N-1}>1$, which is true only for N=2, 3. Phase super-sensitivity cannot be achieved in *any* described nondeterministic scheme for N≥4 [38]. An alternative version of phase super-sensitivity arises if the important physical resource is the number of photons passed through the sample, not the total number of photons consumed. Phase super-sensitivity can then be achieved for all N since in principle time-forward schemes can be heralded with perfect efficiency [17–19]. In a recent work using trapped ions, phase super-resolution was observed for 4, 5, and 6 ions with respective visibilities of 69.8±0.3%, 52.7±0.3%, and 41.9±0.4% [32]. Determining if phase super-sensitivity was achieved requires knowledge of the uncertainty for each data point, which can not be determined from the published data. In the worst case, Eq. 3 shows that phase super-sensitivity was achieved if the overall efficiencies were respectively 51.3±0.4%, 72.0±0.8%, and 95±2% for 4, 5, and 6 ions.

We have used a time-reversal analysis to show that it is not necessary to produce path-entangled states to achieve phase super-resolution. We show that phase super-resolution is possible even in the absence of nonclassical interference; and derive the necessary conditions to claim phase super-sensitivity from phase super-resolution. Using standard laser sources we obtain high-visibility and high-contrast phase super-resolution of up to 6 oscillations per cycle in a six-photon experiment. The improvement in phase resolution is homologous to that achieved in a standard path interferometer driven at a wavelength of 105.5 nm—one-sixth the wavelength of our He:Ne laser. Inverting the roles of state production and measurement is an application of a more general time-reversal analysis technique [29, 30]: given the dramatic improvement demonstrated here, it remains an interesting open question as to which other quantum technologies will benefit from this technique.

This work supported in part by the Australian Research Council and the DTO-funded U.S. Army Research Office Contract W911NF-05-0397. We thank T. C. Ralph and G. J. Milburn for discussions, and H. Wiseman, D. T. Pegg and P. Meredith for helpful MS comments.

[1] H. Lee, et al., *Proc. Sixth Int. Conf. Quant. Comm. Meas. & Comp.* (*Ed*: J. H. Shapiro & O. Hirota) 223-229 (Rinton Press, Princeton, 2002).
[2] V. Giovannetti, et al. *Phys. Rev. Lett.* **96**, 010401 (2006).
[3] B. Yurke, *Phys. Rev. Lett.* **56**, 1515 (1986).
[4] M. Hillery & L. Mlodinow, *Phys. Rev. A* **48**, 1548 (1993).
[5] M. J. Holland & K. Burnett, *Phys. Rev. Lett.* **71**, 1355 (1993).
[6] C. Brif & A. Mann, *Phys. Rev. A* **54**, 4505 (1996).
[7] Z. Y. Ou, *Phys. Rev. A* **55**, 2598 (1997).
[8] R. A. Campos, et al., *Phys. Rev. A* **68**, 023810 (2003).
[9] J. P. Dowling, *Phys. Rev. A* **57**, 4736 (1998).
[10] J. J. Bollinger, et al., *Phys. Rev. A* **54**, R4649 (1996).
[11] A. Boto, et al., *Phys. Rev. Lett.* **85**, 2733 (2000).
[12] M. D'Angelo, et al., *Phys. Rev. Lett.* **87**, 013602 (2001).
[13] E. J. S. Fonseca, et al., *Phys. Rev. A* **63**, 043819 (2001).
[14] M. W. Mitchell, et al., *Nature* **429**, 161 (2004).
[15] P. Walther, et al., *Nature* **429**, 158 (2004).
[16] E. Knill, et al., *Nature* **409**, 46 (2001).
[17] H. Lee, et al., *Phys. Rev. A* **65**, 030101 (2002).
[18] J. Fiurášek, *Phys. Rev. A* **65**, 053818 (2002).
[19] G.J.Pryde & A.G.White, *Phys.Rev.A* **68**, 052315(2003).
[20] X. Zou, et al., quant-ph/0110149 (2001).
[21] P. G. Kwiat, et al., *Phys. Rev. A* **41**, 2910 (1990).
[22] J. G. Rarity, et al., *Phys. Rev. Lett.* **65**, 1348 (1990)
[23] Z. Y. Ou, et al., *Phys. Rev. A* **42**, 2957 (1990).
[24] K. Edamatsu, et al. *Phys. Rev. Lett.* **89**, 213601 (2002).
[25] For example, post-selecting the two-mode state [14], $(|3,0\rangle + |0,3\rangle)/\sqrt{2}$, or four-mode state [15], $(|2,2,0,0\rangle + |0,0,2,2\rangle)/\sqrt{2}$.
[26] M. Reck, et al., *Phys. Rev. Lett.* **73**, 58 (1994).
[27] Y. Aharonov, et al., *Phys. Rev.* **134**, B1410 (1964)
[28] D. T. Pegg, et al. *J. Mod. Opt.* **49**, 913 (2002).
[29] K. L. Pregnell, quant-ph/0508088 (2005).
[30] K.L.Pregnell & D.T.Pegg, *J. Mod. Opt.* **51**, 1613 (2004).
[31] F. W. Sun, et al., *Phys. Rev. A* **73**, 023808 (2006).
[32] D. Leibfried, et al., *Nature* **438**, 639 (2005).
[33] If the photon flux is large enough that the signal is above the electronic noise floor of typical square-law intensity detectors, then our experiment could be further simplified by replacing the photon counters with intensity detectors, and the multiphoton coincidence with electronic signal multiplication. This may not be desirable if the phase shift is significantly affected by high photon flux, e.g. biological specimens, radiation pressure on a mirror.
[34] In a five-photon entanglement experiment (not involving phase super-resolution), the coincidence rate was 0.024 Hz, 175 counts in 2 hours [35].
[35] Z. Zhao, et al., *Nature* **430**, 54 (2004).
[36] S. Bentley & R. Boyd, *Opt. Express* **12**, 5735 (2004).
[37] G. Khoury, et al., *Phys. Rev. Lett.* **96**, 203601 (2006).
[38] The three-photon phase super-resolution experiment [14] did not exhibit phase super-sensitivity as the reported visibility of 42±3% is below $V_{\text{thr}}=1/\sqrt{3}=57.7\%$, even disregarding the low efficiencies of downconversion and photon collection, $\eta \ll 1$.